# New Method to Detect the Transport Scattering Mechanisms of Graphene Carriers

Shuang Tang*[1], Mildred S. Dresselhaus†[2,3]

[1] Department of Materials Science and Engineering, Massachusetts Institute of Technology, Cambridge, MA, 02139-4037, USA; (*tangs@mit.edu; Tel: 617-253-6860; Fax: 617-253-6827)

[2] Department of Electrical Engineering & Computer Science, and [3] Department of Physics, Massachusetts Institute of Technology Cambridge, MA, 02139-4037, USA; († millie@mgm.mit.edu; Tel: 617-253-6864; Fax: 617-253-6827)

**ABSTRACT:** Detecting the carrier scattering mechanisms in a materials system is important for transport related science and engineering. The approaches of fast laser process and electrical conductivity matching were used in previous literature, which do not give accurate information on scattering relaxation time as a function of carrier energy for intrinsic photon-free transport. Graphene is considered as a model system in materials science studying for its simple atomic and electronic structures. Here we have developed a new methodology to detect the scattering relaxation time as a function of carrier energy, which can be used to infer the carrier scattering mechanisms at different temperatures. Our method utilizes the measured values of optimal Seebeck coefficient, for both P-type and N-type materials. This new approach can eliminate the influence from photon-carrier scattering in the fast-laser method, and avoid the over-fitting issue in the electrical conductivity matching method. We have then applied the new approach in the $SiO_2$ substrated graphene system, and discovered that the Dirac carriers are mainly scattered by short-range interactions at 40 K. The scatter strength of long-range Coulomb interaction increases with temperature. At 300 K, the long-range and short-range interactions scatter the Dirac carriers with almost comparable strengths.

Graphene has attracted much research attention since its experimental discovery [1]. Most important properties of graphene are associated with its special band structure of Dirac cones [2] at the *K*- and *K'*-points in the Brillouin zone. The unusual scattering mechanisms of Dirac carriers result in many novel physical phenomena, including the Klein paradox [3-6], the finite minimum conductivity [7-13], etc. Thus, researchers have been very interested in exploring specific carrier scattering mechanisms of Dirac electrons and Dirac holes at different temperatures in various graphene samples [11, 14-26]. So far, various photonic methods have been used by researchers to detect the relaxation time of the graphene carriers [27-32], the results of which can be further used to analyze the carrier scattering mechanisms. However, most reports on carrier scattering relaxation time are fitted values mainly from electrical conductivity measurements [25], which are actually not sufficient to infer the scattering mechanism, especially, when many scattering mechanisms coexist. Further, most reports involve only discussions of the average carrier scattering relaxation time [11, 14-24]. There are barely any reports on detecting the carrier scattering relaxation time as a function of carrier energy or scattering mechanism in transport property studies in the absence of photon interactions. Photon-interaction-free carrier scattering, however, is very important, since the application potential of graphene extends widely beyond only photonic devices [33]. Our present work aims to provide a methodology to study the carrier scattering relaxation time (in the low field regime) by conductivity and Seebeck coefficient measurements, and use this methodology to analyze the scattering mechanisms of carriers at different temperatures. Since the band structure near the Dirac cone of graphene is very simple, it can be used as a model system. The methodology we developed here can be applicable for beyond graphene to general materials systems, especially to materials systems with a simple band structure, including single layer $MoS_2$ and $WS_2$ [34], black phosphorus [35], surface states of topological insulators [36, 37], etc.

In this paper, we first point out that the popularly used Mott relation for Seebeck coefficient calculations cannot capture the values of the optimal Seebeck coefficient of a system. Then we rationalize the "power law" approximation for the carrier scattering relaxation time as well as for the transport distribution function, and provide an approach to treat the cases where many scattering mechanisms coexist. Next, we show that the values of the optimal Seebeck coefficient of a system with both P-type and N-type carriers are primarily determined by the "exponent" in the "power law" approximation of the transport distribution function, and hence, the carrier scattering mechanism is identified only within the limits of this approximation, where both numerical and analytical approaches are discussed. Finally, we use this method to detect the carrier scattering relaxation time and scattering mechanism of graphene on a $SiO_2$ substrate.

## Limitation of the Mott Relation

It is well known that the low field transport of a materials system can be well described by the Boltzmann equation [38] in terms of:

$$\sigma = q^2 I_{[r=0]}, \tag{1}$$

$$S = \frac{k_B}{q} \frac{I_{[r=1]}}{I_{[r=0]}}, \tag{2}$$

$$\kappa_e = \kappa_0 - T\sigma S^2, \tag{3}$$

and

$$\kappa_0 = Tk_B^2 I_{[r=2]}, \quad (4)$$

where

$$I_{[r]} = \int \left(-\frac{\partial f_0}{\partial E}\right) \Xi(E) \left(\frac{E-E_f}{k_B T}\right)^r dE \quad (5)$$

and

$$\Xi(E) = \sum_{\mathbf{k}} \tau(\mathbf{k}) \delta(E-E_{\mathbf{k}}) \mathbf{vv}, \quad (6)$$

where $\sigma$ is the electrical conductivity, $S$ is the Seebeck coefficient, $\kappa_e$ is the electronic thermal conductivity, $T$ is the temperature, $E_f$ is the Fermi level, $f_0$ is the Fermi distribution function, $\tau$ is the relaxation time, and $\mathbf{v}$ is the carrier group velocity. Here $r$ is the order of integration of equation (5), which can be 0, 1 or 2. The function $\Xi$ is called the transport distribution function that is often modeled using an exponential form of $\Xi(E) = \Xi_0 (E/k_B T)^n$, which is well known as a "power law" dependence [38-44]. The transport of graphene, however involves both a chirality and a phase factor shift [25], but these qualities of transport can still be described by the Boltzmann equation using an effective transport distribution function [16, 25].

The Mott relation is widely used in calculations of Seebeck coefficient from the measured electrical conductivity [45-47]. We here point out that the Mott relation only captures the Seebeck coefficient corresponding to a Fermi level that is far away from the energy range where the Seebeck coefficient is optimized; it fails to capture the optimal Seebeck coefficient, as shown in Fig. 1 (a). Consider a more general semiconducting system, where the conduction band valley has a transport distribution function of

$\Xi_C(E) = \Xi_{C,0}(T)(E/k_BT)^{n(T)}$, and the valence band has a transport distribution function of

$\Xi_V(E) = \Xi_{V,0}(T)(E/k_BT)^{n(T)}$. For the purpose of a general illustration, we consider the conduction band and the valence band not to be symmetric to one another, i.e. $\Xi_{C,0} \neq \Xi_{V,0}$. For example, let us consider $5\Xi_{C,0} = \Xi_{V,0}$, and assume that the band gap is 10 $k_BT$. The discrepancy between the Seebeck coefficient from the Boltzmann equation and the Seebeck coefficient from the Mott relation, as shown in Fig. 1 (a), can be explained by the difference between the Mott relation and equation (2). From the Boltzmann equation given by equation (2), we have the real Seebeck coefficient for a single valley as,

$$S = \frac{k_B}{e} \frac{\int \left(-\frac{\partial f_0}{\partial E}\right) \Xi(E) \left(\frac{E - E_f}{k_BT}\right) dE}{\int \left(-\frac{\partial f_0}{\partial E}\right) \Xi(E) dE}, \tag{7}$$

while the Mott relation [45-47], i.e.

$$S_{Mott}(E_f) = -\frac{\pi^2 k_B^2 T}{3e\sigma} \left(\frac{\partial \sigma}{\partial E}\right)_{E_f}, \tag{8}$$

gives the approximated Seebeck coefficient $S_{Mott}$ as

$$S_{Mott} = \frac{k_B}{e} \frac{\int \left(-\frac{\partial f_0}{\partial E}\right) \Xi(E) \frac{\pi^2}{3} \tanh\left(\frac{E - E_f}{2k_BT}\right) dE}{\int \left(-\frac{\partial f_0}{\partial E}\right) \Xi(E) dE}. \tag{9}$$

In essence, the Mott relation is using the hyperbolic tangent relation $(\pi^2/3)\tanh[(E-E_f)/2k_BT]$ to approximate the linear relation $(E-E_f)/k_BT$, which can make a significant difference with Seebeck coefficient determined over the Fermi level range where the optimal Seebeck coefficient occurs. Hence, we see that the Mott relation used widely in the thermoelectrics literature is only valid to model the Seebeck coefficient far away from the conditions where the optimal Seebeck coefficient occurs. However, the real values of optimal Seebeck coefficient for a materials system can provide us with a tool to detect the carrier scattering mechanism, as will be exhibited in the following discussion.

**The Power Law Theorem for Multiple Scattering Mechanisms**

Before developing the tool for detecting the carrier scattering mechanism, we rationalize the "power law" approximation for the carrier scattering relaxation time as well as the transport distribution function, and provide an approach to treat the cases where multiple scattering mechanisms coexist. Under the relaxation time approximation, the functional form of the relaxation time $\tau(E)$ is fundamentally determined by the micro-mechanisms of the electron and hole scattering. For example, in graphene, the acoustic phonon scattering relaxation time can be written as $\tau_{AP}(E)^{-1} = \left(4v^2\hbar^3\rho v_s^2\right)^{-1}\left(U_A^2 k_B T\right)E$, where $U_A$ is the acoustic deformation potential, $\rho$ is the mass density, $v$ is the group velocity of Dirac fermions, and $v_s$ is the velocity of sound [16, 25, 48]. For short-range disorder scattering, $\tau_{SD}(E)^{-1} = \left(4\hbar^3 v^2\right)^{-1}\left(n_d U_d^2\right)E$, where $U_0$ is the short-range disorder interaction potential, $n_d$ is the short-range disorder density [16, 25, 48]. For long-range Coulomb interaction scattering, $\tau_C(E)^{-1} = (u_0^2/\hbar)E^{-1}$, where $u_0$ is an effective charge parameter, which takes into consideration of the Coulomb interaction and the screening effect [16, 25, 48]. For the vacancy scattering mechanism, $\tau_V(E)^{-1} = \left(4\hbar\delta_k^2 v^2 n_v\right)^{-1}E$, where $n_v$ is the vacancy density and $\delta_k$ is the phase factor shift [16, 25, 48]. For electron-electron scattering, the form of $\tau$ is more complicated, but $\tau$ can be in general captured by a polynomial of the carrier energy [48]. The overall relaxation time follows the Matthiessen's rule [49] $\tau^{-1} = \sum_i \tau_i^{-1}$, where $i$ stands for different carrier scattering mechanisms. Thus, we can see that $\tau$ can be modeled as a polynomial of carrier energy too, as $\tau = \sum_N c_N (E/k_B T)^{s_N}$, where $N$ is used as a label for the different scattering mechanisms, $s_N$ can be positive, negative or zero, and $c_N$ are functions of temperature. A further simplification can be made as

$$\tau(E) = \tau_0(T) \left( \frac{E}{k_B T} \right)^{s(T)}, \quad (10)$$

where $s(T)$ is a certain average of $s_N$, e.g. without loss of generality, we can take the form of $s(T)$ to be $s(T) = \langle s_N \rangle_T = \sum_N s_N e^{c_N} / \sum_N e^{c_N}$. This explanation is consistent with the original literature proposing the "power law" approximation of $\tau$. [38-44] Here $\tau_0(T)$ is a function of temperature, which does not matter in evaluating Seebeck coefficient, and can be determined by the electrical conductivity measurement. We know that [50] $\Xi(E) = \tau(E) D(E) \langle v^2 \rangle_E$, where $\langle . \rangle_E$ stands for the mean value on the constant energy surface. Here $\langle v^2 \rangle_E \propto E^j$ and $j=0$ ($j=1$) for a linear (parabolic) band. Thus, we can write

$$\Xi(E) = \Xi_0(T) \left( \frac{E}{k_B T} \right)^{n(T)}, \quad (11)$$

by putting $\tau(E)$, $D(E)$ and $\langle v^2 \rangle_E$ together with $n=s+l+j$, where the temperature dependence of $\Xi_0$ and $n$ both come from the relaxation time $\tau(E)$.

## Relate the Optimal Seebeck Coefficient to Scattering Mechanisms

Next, we show that the values of the optimal Seebeck coefficient for any P-type or N-type system are primarily determined by the "exponent" in the "power law" approximation of the transport distribution function, i.e. by $n$ in equation (11), and hence, by the carrier scattering mechanism, which is reflected by $s$ in equation (10). From equation (2) and (5) we obtain the Seebeck coefficient for a single band valley as,

$$S(E_f) = \frac{1}{qT} \left( \frac{\int_0^\infty \left(\frac{E}{k_B T}\right)^{n(T)+1} \left(-\frac{\partial f}{\partial E}\right)_{E_f} dE}{\int_0^\infty \left(\frac{E}{k_B T}\right)^{n(T)} \left(-\frac{\partial f}{\partial E}\right)_{E_f} dE} - E_f \right). \qquad (12)$$

In order to obtain the relation between the Seebeck coefficient and $n(T)$, we first introduce the hypothesis that $n(T)$ for a specific band valley can be tuned artificially through a certain manner, i.e. the exponent $s$ in equation (10) can be tuned by artificially changing the scattering mechanism. Thus, we can take the derivative of $S(E_f)$ in equation (1) as,

$$\frac{dS(E_f)}{dn(T)} = \frac{1}{qT}\left[n(T)+1-n(T)\frac{\int_0^\infty E^{n(T)+1}\left(-\frac{\partial f}{\partial E}\right)_{E_f} dE \int_0^\infty E^{n(T)-1}\left(-\frac{\partial f}{\partial E}\right)_{E_f} dE}{(\int_0^\infty E^{n(T)}\left(-\frac{\partial f}{\partial E}\right)_{E_f} dE)^2}\right] \approx \frac{1}{qT}. \qquad (13)$$

Therefore, we have the Seebeck coefficient to be a linear function of $n(T)$ to lowest order,

$$S(E_f) \approx \frac{n(T)}{qT} + S_{n=0}(E_f), \qquad (14)$$

where the next order is not as important compared to this linear relation, as discussed below.

We consider again the system in Fig. 1(a), where the optimal Seebeck coefficient for different $n(T)$ in Fig. 1 (a) is exhibited in Fig. 1 (b) for both P-type and N-type thermoelectric materials. The results of Fig. 1(b) clearly demonstrate that the optimal Seebeck coefficient for both P-type and N-type materials changes linearly with $n(T)$. On the one hand, these results imply that if the exponent $n(T)$ in equation (11) or the exponent $s(T)$ in equation (10) is improperly chosen, the optimal Seebeck coefficient will be mis-evaluated, no matter how well the improperly chosen equation (11) or equation (10) can be used to fit the electrical conductivity. One the other hand, the results also imply that we can take advantage of the fact that the values of the optimal Seebeck coefficient are monotonically determined by the carrier scattering mechanism, which is reflected by the exponent $s$ in equation (10). This tells us how we can measure the optimal values of the Seebeck coefficient and then use this information to detect the carrier scattering mechanism of a system at different temperatures.

## Application to Graphene on a SiO$_2$ Substrate

Finally, we use the above-derived methodology to obtain the Dirac carrier scattering relaxation time for different carrier energies in graphene at different temperatures, and analyze the implied scattering mechanisms. The measurements of the Seebeck coefficient of graphite were made many decades ago [51, 52]. With the help of nanotechnology in device fabrication, the measurement of the Seebeck coefficient as a function of Fermi level is now available [53], which provides us with a convenient system for the application of our approach for studying the transport process and scattering mechanisms. The graphene system has two isotropic Dirac cones with their apexes at around the Fermi level in the Brillouin zone. The transport distribution for such a two-dimensional linear band valley is $\Xi(E) = 2(E/2\pi\hbar^2)\tau(E)$.[25] Thus, if we can get $n(T)$ from experimental measurements at different temperatures, we can find the specific form of $\tau = \tau_0(T)(E/k_BT)^{s(T)}$ from equation (11). The Seebeck coefficient from the Dirac cones can be calculated as,

$$S = \frac{k_B}{e} \frac{\int_{\text{Conduction Band}} \left(-\frac{\partial f_0}{\partial E}\right) \Xi_C(E) \left(\frac{E-E_f}{k_BT}\right) dE - \int_{\text{Valence Band}} \left(-\frac{\partial f_0}{\partial E}\right) \Xi_V(E) \left(\frac{E-E_f}{k_BT}\right) dE}{\int_{\text{Conduction Band}} \left(-\frac{\partial f_0}{\partial E}\right) \Xi_C(E) dE + \int_{\text{Valence Band}} \left(-\frac{\partial f_0}{\partial E}\right) \Xi_V(E) dE},$$

(15)

where $\Xi_C(E) = \Xi_{C,0}(T)(E/k_BT)^{n(T)}$ and $\Xi_V(E) = \Xi_{V,0}(T)(E/k_BT)^{n(T)}$. The ratio $\Xi_{V,0}/\Xi_{C,0}$ characterizes the asymmetry between the conduction band and the valence band at different temperatures. We also see that the asymmetry ratio of the transport distribution functions $\Xi_{V,0}/\Xi_{C,0}$ is equal to the asymmetry ratio of the relaxation time $\tau_{V,0}/\tau_{C,0}$ for the carriers associated with the Dirac cones of graphene. Based on our results above from equation (14) and Fig. 1 (b), we see that the optimal Seebeck coefficient for each single band depends only on $n(T)$. However, we have to consider both the conduction band and the valence band here, so the optimal Seebeck coefficient also depends on $\tau_{V,0}/\tau_{C,0}$. We can first calculate a map of the optimal Seebeck coefficient as a function of $n(T)$ and the asymmetry ratio $\tau_{V,0}/\tau_{C,0}$ for both the P- and the N-type, as shown in Fig. 2 (a)-(c). Then by matching the measured optimal Seebeck coefficient to this map, we can obtain a single solution corresponding to the measured $n(T)$ and $\tau_{V,0}/\tau_{C,0}$ at each different temperature. The optimal Seebeck coefficient at different temperatures for the $SiO_2$ subsrated graphene has been measured by Ref. Zuev et al. [53] through a gated thermoelectric device by changing the Fermi level, and their results are summarized in Table I. Based on Table I, we have calculated $n(T)$ and $\tau_{C,0}/\tau_{V,0}$ at each different temperature, as shown Fig. 2 (d).

From Fig. 2 (d), we see that the "exponent" $n(T)$ changes significantly with temperature, which implies that the carrier scattering mechanism is very temperature sensitive. At the low temperature end, the exponent $n(T) \to 0$ and the exponent $s(T) \to -1$, which implies that $\tau^{-1} \propto D(E)$. At the high temperature end, $n(T) \to 1$ and $s(T) \to 0$, which implies that $\tau^{-1}$ is a constant over a different range of carrier energies. The measured $s$ values then give us important information about the important carrier scattering mechanisms, at the various temperatures, since the $E$-dependence of $\tau^{-1}$ in equation (10) varies with scattering mechanisms.

For the low temperature range, $\tau \propto 1/D(E) \propto E^{-1}$, which implies that the dominant scattering mechanisms should be acoustic phonon scattering, short-range disorder scattering, and vacancy scattering. We know that in the cryogenic temperature range, the acoustic phonon modes are frozen and could not contribute significantly to the carrier scattering. Thus, we know that in the cryogenic temperature range, for the device configuration utilized in Ref. [53], where the graphene is sitting on a $SiO_2$ substrate, the scattering of the Dirac carriers, comes mainly from the short-range interaction of point defects and vacancies that are intrinsically formed within the graphene sheet [54]. This result is consistent with the intuition that the minimum resistance of a material comes from the intrinsic defects formed within the material system itself.

For the temperature range up to 300 K, $\tau^{-1}$ becomes almost independent of carrier energy, which implies that there is (are) a (some) scattering mechanism (mechanisms) that has (have) a larger *n* in equation (10) that becomes more and more important as the temperature increases. In the system where graphene is on a $SiO_2$ substrate, the most important scattering mechanism, which has a value of *n* that is larger than 1, is the long-range Coulomb interaction scattering, where $\tau \propto E$. This gives us important information that the interaction between the graphene sheet and the charged impurities embedded in the surface of the $SiO_2$ substrate will increase with temperature. This temperature dependence may be due to many causes. The strength of the long-range Coulomb interaction depends exponentially on the distance between the graphene sheet and the surface of the $SiO_2$ substrate. Increasing the temperature leads to an increased amplitude and strength of vibrations of the graphene sheet in the direction perpendicular to the $SiO_2$ substrate, which reduces the average effective interaction distance between the graphene and the substrate surface [55]. Further, the increased temperature will cause dipole vibrations in the optical phonon modes of $SiO_2$, which will further stimulate the long-range Coulomb interaction [33, 56, 57]. Based on the above results, we can predict that increasing the strength of the long-range Coulomb interaction from charged impurities will increase the optimal Seebeck coefficient for both P-type and N-type graphene.

Another important trend we see from Fig. 2 (d) is that the asymmetry ratio $\tau_{V,0}/\tau_{C,0}$ increases with temperature, where the conduction band and the valence band are close to being symmetric with one another at a temperature as low as 40 K. This is consistent with the report that electrons and holes are asymmetric in the transport in graphene related systems [58], though they are symmetric in the dispersion relation. Furthermore, based on the results of Fig. 2 (d), we have calculated the electrical conductivity at $T$=300 K, which are compared with the experimentally measured values [53] in Fig. 3. The relaxation times for the conduction and the valence band are calculated to be $\tau_{V,0} = 3.8 \times 10^{-14}$ sec and $\tau_{C,0} = 2.0 \times 10^{-14}$ sec, respectively. Our calculated electrical conductivity curve is in good agreement with the experiment. We also calculated the electronic thermal conductivity for the Dirac fermions in graphene, as shown in Fig. 3, which is consistent with previously estimated values.

## Discussion

We have pointed out drawbacks to two popular presently used models for the Seebeck coefficient, including Mott relation and the simple constant relaxation time approximation. By taking advantage of the property that we have shown, namely that the values of the optimal Seebeck coefficient are monotonically determined by the carrier scattering mechanism, we have developed a method for detecting the carrier scattering relaxation time corresponding to different carrier energy ranges for the graphene Dirac fermions for electrons and holes, by measuring the optimal values of the Seebeck coefficient.

We have found that at cryogenic temperatures, the Dirac carriers in graphene sitting on a $SiO_2$ substrate are mainly scattered by short-range interactions with the point defects and the vacancies that are formed intrinsically within the graphene sheet. As the temperature increases, the long-range Coulomb interaction, coming from the charged impurities and optical phonon modes on the surface of the $SiO_2$ substrate, become more and more important. At room temperature, the long-range interaction and the short-range interaction both scatter the Dirac carriers with almost comparable strengths. Based on these results, we have also predicted that the optimal Seebeck coefficient for both P-type and N-type graphene can be increased by increasing the strength of the long-range Coulomb interaction coming from charged impurities.

Our method for analyzing the carrier scattering mechanism uses not only measurements of the electrical conductivity, but also of the Seebeck coefficient to deduce the carrier scattering mechanism. This methodology has provided an approach to detect the carrier scattering relaxation time as a function of carrier energy, and not only the average relaxation time for all carriers. The methodology has also provided an approach to infer the relaxation time when multiple scattering mechanisms coexist. By taking the optimal values of the Seebeck coefficient into consideration, we can avoid over-fitting the data, as can be found in some cases in the literature, where there is not a single clear solution for the carrier scattering relaxation time, due to the over-fitting. Our method can be also applicable to more general materials systems, as discussed in the present work, especially to materials systems with a simple band structure, including single-layered $MoS_2$ and $WS_2$, black phosphorous, and surface states of topological insulators.

## Acknowledgements

We acknowledge the support from AFOSR MURI Grant Number FA9550-10-1-0533, Subaward 60028687. The authors thank Dr. Albert Liao for valuable discussions.

## Author Contributions

S. T. conceived the idea, made the calculations and wrote the paper. M. S. D. provided the research platform and funding, and helped in writing the paper.

## Competing financial interests

The authors declare no competing financial interests.

## Figures Legends

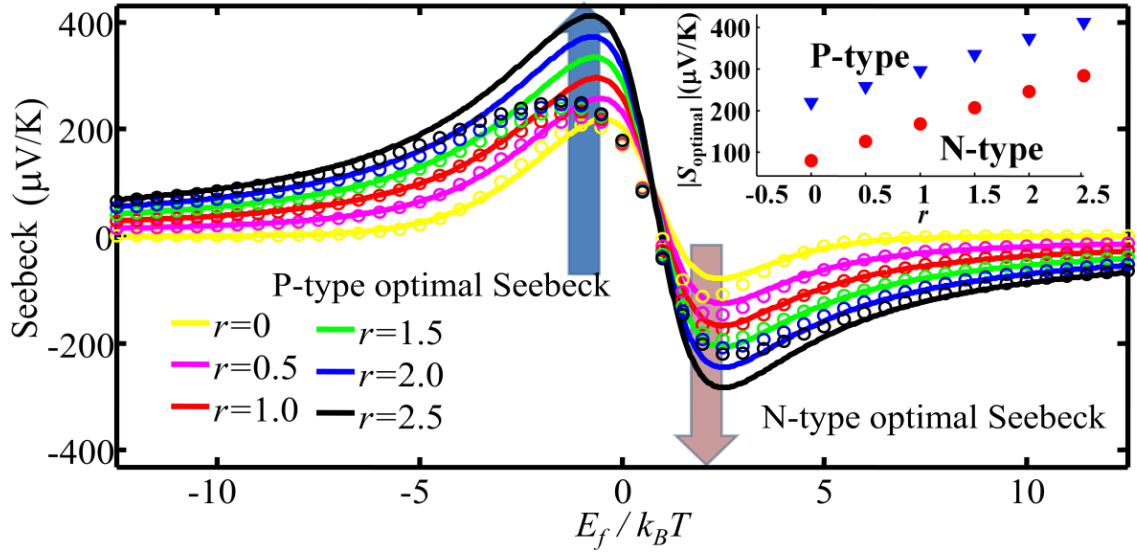

**Figure 1:** (a) Solid lines show the Seebeck coefficient as a function of Fermi Level for a simple system with $5\Xi_{C,0} = \Xi_{V,0}$ and $E_g=10\ k_BT$, when different values of $n$ are taken. The circle lines are the Seebeck coefficient as approximated by the Mott relation (equation (9)), which fails to capture the optimal values of the Seebeck coefficient. (b) The absolute value of the optimal Seebeck coefficient is clearly shown to increases with $n$ in a linear manner in the ranges of Fermi level and temperature we discussed in the present work.

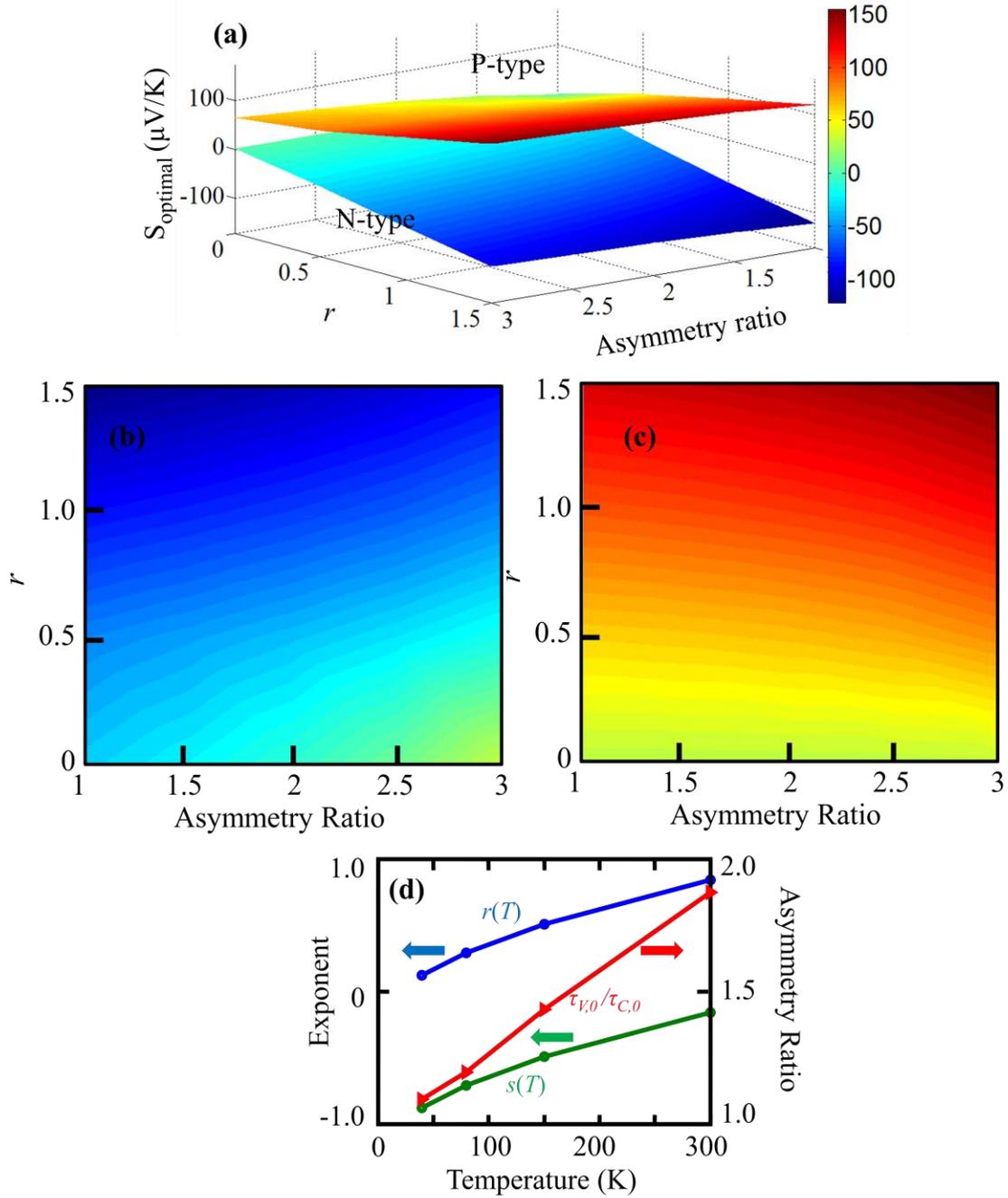

**Figure 2:** Relation of the transport distribution function, relaxation time and thermoelectric properties of the Dirac cone in graphene. (a) The plots for the optimal Seebeck coefficient of a graphene Dirac cone are presented for both N-type and P-type, as a function of $n(T)$ and also for the asymmetry ratio of scattering relaxation times

$\tau_{T,0}/\tau_{C,0}$. To further clarify the two maps of the optimal Seebeck coefficient in (a), we show maps of the optimal Seebeck coefficient vs. the exponent *r* and the asymmetry ratio $\tau_{V,0}/\tau_{C,0}$ for N-type and P-type carriers in (b) and (c), respectively. (d) The calculated *n*(*T*) (blue), *s*(*T*) (green) and asymmetry ratio $\tau_{V,0}/\tau_{C,0}$ (red) of carrier relaxation times for the Dirac cone in graphene.

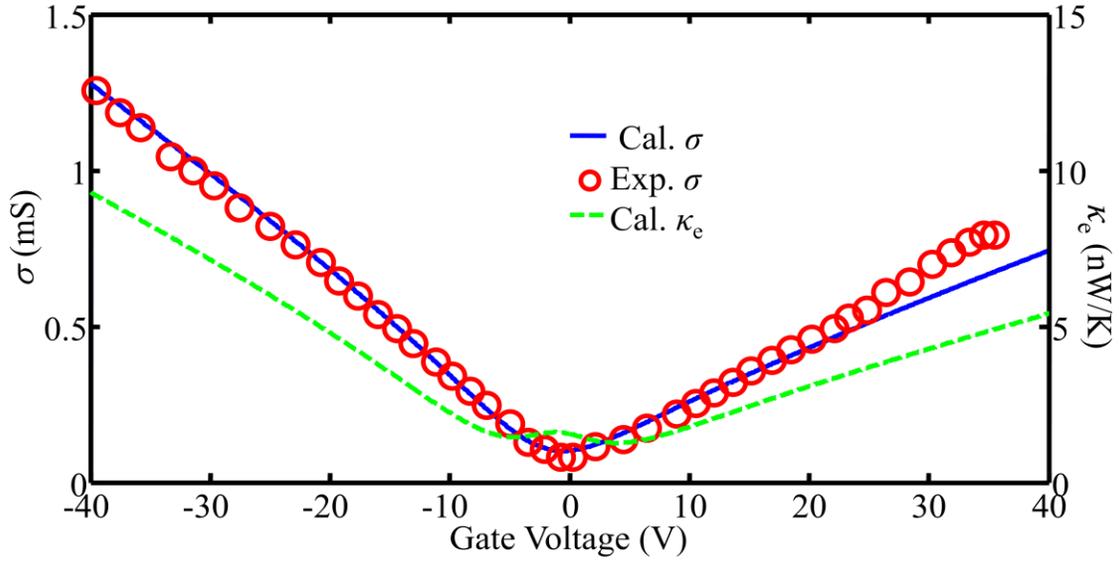

**Figure 3**: The calculated electrical conductivity (blue solid line) for graphene associated with two isotropic Dirac cones at 300 K is compared with the experimentally measured values [53] (red circles). The associated electronic contribution to the thermal conductivity is also calculated ( green dashed line). The asymmetry ratio in the scattering time, namely $\tau_{V,0}/\tau_{C,0}$=1.8, can be seen from the electrical and thermal conductivity plots.

**Table I:** Measured Optimal Seebeck Coefficient [53] for the Dirac Cone in Graphene on a SiO$_2$ substrate

| Temperature (K) | 300 | 150 | 80 | 40 |
|---|---|---|---|---|
| Optimal P-type Seebeck coefficient [53] (μV/K) | 92.52 | 57.94 | 33.64 | 14.95 |
| Optimal N-type Seebeck coefficient [53] (μV/K) | -59.81 | -39.25 | -24.30 | -10.28 |